\documentclass[10pt,twocolumn]{article} 

\usepackage{amsfonts,amsmath,amssymb,mathrsfs}
\usepackage{color,epsfig,graphicx,graphics}
\usepackage{hyperref}
\usepackage{braket}

\usepackage{natbib}
\def\mbb{\mathbb}
\newtheorem{theorem}{Theorem}[section]

\newtheorem{corollary}{Corollary}[section]

\newtheorem{definition}{Definition}[section]

\newtheorem{lemma}{Lemma}[section]

\newtheorem{remark}{Remark}[section]
\numberwithin{equation}{section}

\newcommand{\beq}{\begin{equation}}
\newcommand{\eeq}{\end{equation}}
\newcommand{\beqm}{\begin{equation*}}
\newcommand{\eeqm}{\end{equation*}}
\newcommand{\beqn}{\begin{eqnarray}}
\newcommand{\eeqn}{\end{eqnarray}}
\newcommand{\beqnm}{\begin{eqnarray*}}
\newcommand{\eeqnm}{\end{eqnarray*}}
\newcommand{\bea}{\begin{align}}
\newcommand{\eea}{\end{align}}
\newcommand{\beam}{\begin{align*}}
\newcommand{\eeam}{\end{align*}}

\begin{document}

\title{Structural Decomposition for Quantum Two-level Systems}

\author{G. F. Zhang\thanks{Corresponding author. Department of Applied Mathematics, Hong Kong Polytechnic University, Hong Kong, China
  (guofeng.zhang@polyu.edu.hk, \url{https://www.polyu.edu.hk/ama/profile/gfzhang/}).}
\and Ian~R.~Petersen\thanks{Research School of Electrical, Energy and Materials Engineering, The Australian National University, Canberra ACT 2601, Australia.
  (i.r.petersen@gmail.com).}
}

\maketitle

\begin{abstract}
An input-output model of a two-level quantum system in the Heisenberg picture is of bilinear form with constant system matrices, which allows the introduction of the concepts of controllability and observability in analogy with those of quantum linear systems. By means of the notions of controllability and observability, coordinate transformations, which are rotation matrices, can be constructed explicitly that transform an input-output model to a new one.  The new input-output model enables us to investigate many interesting properties of the two-level quantum system, such as steady-state solutions to the Lindblad master equation, quantum decoherence-free (DF) subspaces, quantum non-demolition (QND) variables, and the realization of quantum back-action evading (BAE) measurements. The physical system in (Wang, J. \& Wiseman, H. M. (2001), Feedback-stabilization of an arbitrary pure state of a two-level atom, Physical Review A 64(6), 063810) is re-studied to illustrate the results presented in this paper.
\end{abstract}


\section{Introduction}\label{sec:intro}

The last few decades have witnessed the fast growth of theoretical advances and experimental demonstrations of quantum control as it is an essential ingredient of quantum information technologies including quantum communication, quantum computation, quantum cryptography, quantum ultra-precision metrology, and nano-electronics. One of the features of quantum control which is drastically different from classical control is that measurement unavoidably disturbs the system of interest in a stochastic manner. To bypass this difficulty associated with measurement-based feedback control, coherent feedback control has been proposed and extensively studied in the quantum control community. An interesting comparison between measurement-based feedback control and coherent feedback control can be found in \cite{NY14b}. Besides these two closed-loop quantum feedback control methods, open-loop quantum  control is also widely used for controlling quantum systems. Among various open-loop control approaches, coherent control has been proven very effective  for controlling finite-level quantum systems. Given a quantum system immersed in a dissipative environment, the non-unitary evolution of the reduced system density operator is governed by the  Lindblad master equation, parametrized by the system Hamiltonian and the Lindblad coupling operator. The system Hamiltonian may consist of two parts:  a free Hamiltonian and a controlled Hamiltonian. The controlled Hamiltonian can be manipulated by an external field (e.g., a laser field) which serves as control signal. Coherent control of quantum finite-level systems concerns how to engineer the controlled Hamiltonian so that the system density operator can be steered in a desired manner, see. e.g.,  \cite{DD01},  \cite{KGR02}, \cite{DH08}, \cite{BCS09},  \cite{LK09}, \cite{BBR10}, \cite{HY13}, \cite{RBR18} and references therein. In these studies, Lie algebraic methods are the major mathematical tools being used,  \cite{MSW68}, \cite{MW98}, \cite{WP03},  \cite{NJ85}, based on which notions  such as reachability and controllability have been proposed and investigated intensively.

In this paper, we study the structure of open quantum two-level systems from a perspective different from those in the literature of coherent control as discussed in the preceding paragraph. Instead of the  Lindblad master equation for the reduced system density operator, we focus on an input-output model in the Heisenberg picture, \cite{DMP+12}, \cite{DMP+16}. Given a set of parameters for the system Hamiltonian and the  system-environment coupling, the resulting input-output model is of a bilinear form with constant system matrices; see Eqs. \eqref{sys1_a}-\eqref{sys1_b}   and \eqref{A_0}-\eqref{A} in Section \ref{sec:setup} below. There are there types of system matrices:  one for system variables which is denoted by $A$, one involved in the coupling between the system and the input fields, which is denoted by $B$, and one involved in the coupling between the system and the output fields, which is denoted by $C$. Also there is an additional vector denoted by $A_0$. Thus, by means of the controllability matrix of the matrix pair $(A,B)$ and the observability matrix of the matrix pair $(A,C)$, analogous to linear system theory  (\cite{ZGPG18}),  a real orthogonal matrix $T$ could be constructed, which transforms the original input-output model to a new one. The transformed system is still a quantum system in the sense that the fundamental commutation relations among the system variables have to be preserved. This imposes constraints  on the coordinate transformation matrix $T$. It turns out that $T$ must be a 3-by-3 rotation matrix; see Theorem \ref{thm:SO3_aug13}.  The structural decomposition is constructed in Theorems \ref{thm:c2=0} and \ref{thm:general}.  This  structural decomposition enables the straightforward characterization of the existence of stationary solutions to the  Lindblad master equation \eqref{eq:master}, see item (1) of Remark \ref{rem:general}. Moreover, the structural decomposition shows in a transparent way under what conditions the two-level quantum system has a quantum DF  subspace, QND  variables and realizes quantum BAE measurement, see Remarks \ref{rem:special} and  \ref{rem:general} for detail. The physical system in \cite{WW01} is re-studied in Section \ref{sec:example} to illustrate the results presented in this paper.

The rest of the paper is organized as follows. In Section \ref{sec:pre}, we present two properties of skew-symmetric matrices in $\mbb{R}^{3\times 3}$. In Section \ref{sec:setup}, we introduce the input-output model of open two-level systems. In Section  \ref{sec:special}, we investigate a special case of the system decomposition and move on to the general case in Section \ref{sec:general}. In Section \ref{sec:example}, we use a physical example to illustrate the main results of the paper. Section \ref{sec:con} concludes the paper.

{\it Notation.} $\imath = \sqrt{-1}$ is the imaginary unit. $x^\ast$ denotes the complex conjugate of a complex number $x$ or the adjoint of an operator $x$. Given a matrix $X$, let $\mathrm{Re}(X)$ and $\mathrm{Im}(X) $  denote its real part and imaginary part, and $\mathrm{Ker}\left( X\right)$ and $\mathrm{Range}\left(X\right)$ denote the null space and the range, respectively. 
 For a matrix $X=[x_{ij}]$ with number or operator entries,  denote the transpose $X^{\top}=[x_{ji}]$ and conjugate transpose  $
X^{\dag }=[x_{ji}^\ast]$.    Given two column vectors of operators $X$ and $Y$, their commutator is defined to be
\begin{equation}
[X,\ Y^\top] \triangleq XY^\top - (YX^\top)^\top.
\end{equation}
Finally,  ${\rm SO}(3)$ denotes the 3D rotation group (alternatively called the special orthogonal group in three dimensions) \cite[Section 1.2]{NJ85}.

\section{Skew-symmetric matrices in $\mbb{R}^{3\times 3}$}\label{sec:pre}

Given a {\em real} vector $\beta$ of dimension 3, no matter whether  it is a row or column,  define a skew-symmetric matrix
\[
\Theta (\beta)
\triangleq
 \left[ 
\begin{array}{ccc}
0 & \beta _{3} & -\beta _2 \\ 
-\beta _{3} & 0 & \beta _1 \\ 
\beta _2 & -\beta _1 & 0
\end{array}
\right],
\]
where $\beta_1,\beta_2,\beta_3$ are entries of the vector $\beta$. Properties of the skew-symmetric matrix $\Theta (\beta)$ can be found in  \cite[Lemma 1]{DMP+12}. Two more properties are given  below.
\begin{lemma}\label{thm:ian}
An arbitrary matrix $M = [m_1\ m_2 \ m_3]\in \mbb{R}^{3\times 3}$ satisfies
\[
m_1^\top \Theta(m_2) m_3 = -{\rm det}(M),
\]
where ${\rm det}(M)$ denotes the determinant of the matrix $M$.
\end{lemma}

\begin{lemma}\label{thm:SO3}
An orthogonal matrix $T\in \mbb{R}^{3\times 3}$ satisfies  
\begin{equation}  \label{T_CCR}
\Theta(T^\top\beta) = T^\top \Theta(\beta) T, ~~~ \forall \beta \in\Bbb{R}^3
\end{equation}
 if and only if  $T \in {\rm SO}(3)$.
\end{lemma}

These two lemmas can be established by straightforward algebraic manipulations, and hence their proofs are omitted.

\section{An input-output model of two-level systems}\label{sec:setup}


A computational basis of a quantum two-level system consists of two orthonormal basis ket vectors, say  $\ket{0}$ and $\ket{1}$. Let $\bra{0}$ and $\bra{1}$ be their dual basis bra vectors, respectively.  Define operators $\sigma_- = \ket{0}\bra{1}$ and $\sigma_+ = \ket{1}\bra{0}$. Then the Pauli matrices are $\sigma_1 = \sigma_+ + \sigma_-$, $\sigma_2 = \imath(\sigma_+ - \sigma_-)$, and  $\sigma_3 = \sigma_-\sigma_+ -\sigma_+ \sigma_-$.  Stacking these traceless Hermitian operators in a column vector we define $X\triangleq [\sigma _1 \ \ \sigma _2 \ \ \sigma _3]^\top$.  The commutator for $X$ is 
\[
[X, X^\top] 
=2\imath  \left[
\begin{array}{ccc}
0                & \sigma_3  & -\sigma_2 \\
-\sigma_3  & 0               & \sigma_1\\
\sigma_2   & -\sigma_1  &0         
\end{array}
\right],
\]
which we {\it informally} re-write as
\begin{equation}\label{eq:aug24_X}
[X, X^\top] = 2\imath \Theta(X).
\end{equation}
The system Hamiltonian of a two-level system can be described by $H=\alpha X$, where $\alpha =
[\alpha _1  \ \  \alpha _2  \ \  \alpha _{3}]$ is a real row vector. The coupling between the two-level system and its surroundings can be described by a Lindblad operator  $L=\Gamma X$, where $\Gamma $ is a  row vector in the complex domain.  Given $\Gamma$, define two real row vectors $c_1, c_2$ and two real matrices $B_1, B_2$.
\begin{subequations}
\begin{eqnarray}
&& c_1 \triangleq 2\mathrm{Re}(\Gamma),  \ \ c_2 \triangleq 2\mathrm{Im}(\Gamma),
\label{c1}  
\\
&&B_1  \triangleq \Theta (c_{2}) , \ \ B_2  \triangleq -\Theta (c_{1}).
\end{eqnarray}
\end{subequations}
 The dynamics of  a two-level system can be described by an input-output model in the Heisenberg picture (\cite[Eqs. (1.4)-(1.6)]{DMP+16}, \cite[Lemma 2]{DMP+12}),
\begin{subequations}
\begin{eqnarray}
dX 
&=&
A_{0}dt+AXdt+B\left[ 
\begin{array}{c}
XdW_1 \\ 
XdW_{2}
\end{array}
\right] , 
\label{sys1_a} 
\\
\left[ 
\begin{array}{c}
dY_1 \\ 
dY_{2}
\end{array}
\right]  
&=&
CXdt+\left[ 
\begin{array}{c}
dW_1 \\ 
dW_{2}
\end{array}
\right] ,   \ \ t\geq 0,
\label{sys1_b} 
\end{eqnarray}
\end{subequations}
where $W_{1}$ and $W_{2}$ are two quadrature operators of the input field, and the system matrices are
\begin{subequations}
\begin{eqnarray}
&&A_{0}=
\Theta(c_2)c_1^\top, \ B=[ 
\begin{array}{cc}
B_1 & B_{2}
\end{array}
] ,  \ C= \left[ 
\begin{array}{c}
c_1 \\ 
c_{2}
\end{array}
\right] 
\label{A_0}
\\
&&A =-2\Theta (\alpha)-\frac{1}{2}BB^{\top}.
\label{A}
\end{eqnarray}
\end{subequations}
The interested reader may refer to \cite[Chapter 5.3]{GZ00} for an excellent introduction to input-output models of quantum Markovian systems.

In the coherent control literature, the dynamics of the reduced system density operator $\rho$ of a two-level system is characterized by the Lindblad master equation 
\beq \label{eq:master}
\dot{\rho} = -\imath [H, \rho] + \mathcal{L}_L(\rho),
\eeq
 where  $\mathcal{L}_L(\rho) =L\rho L^\ast - \frac{1}{2} L^\ast L \rho -\frac{1}{2}\rho L^\ast L$.  $\rho(t)$ is often parameterized by Bloch parameters, i.e.,  $\rho(t) =\frac{1}{2}(I+a_1(t)  \sigma_1 +a_2(t)  \sigma_2+a_3(t)  \sigma_3)$,  where $a_1(t), a_2(t), a_3(t)\in \mathbb{R}$ satisfy $a_1^2+a_2^2+a_3^2\leq 1$.  $\rho$ is  a pure state if any only if $a_1^2+a_2^2+a_3^2=1$. Denote the Bloch vector by a real column vector $a(t) = [a_1(t) \ a_2(t) \ a_3(t)]^\top$. Then it can be verified that
\beq \label{eq:jun27_a_2} 
\dot{a} = A a + A_0
\eeq
with $A$ and $A_0$ given in \eqref{A_0}-\eqref{A}. Eq. \eqref{eq:jun27_a_2} is usually called the Bloch equation. In the literature of quantum computation, the subspace of states $\tilde{\rho}$ satisfying $\mathcal{L}_L (\tilde{\rho})=0$ is called a DF  subspace  (\cite{LCW98}). In this case, $\tilde{\rho}$ evolves unitarily, and the corresponding Bloch equation \eqref{eq:jun27_a_2}  reduces to $\dot{a} = -2\Theta(\alpha) a$.

\begin{remark}
Several other forms of the Bloch equation \eqref{eq:jun27_a_2}  have been proposed in the literature, see, e.g., \cite[Eq. (2)]{HY13}, \cite[Eq. (4)]{RBR16}, and \cite[Eq. (6)]{SW10}, where the starting point is a Lindblad master equation \eqref{eq:master}.  Here we see that the Bloch equation \eqref{eq:jun27_a_2} is a natural consequence of the input-output model \eqref{sys1_a}-\eqref{sys1_b}. 
\end{remark}

In addition to DF subspaces, QND  variables, and quantum BAE measurements  are also important concepts in quantum information science and quantum measurement theory. An observable $F$ is called a continuous-time QND variable if the commutator $
[F(t_1),F(t_2)]=0$ for all  $t_1,t_2 \in \mathbb{R}^+$, \cite[Eq. (2)]{BVT80}, \cite[Eq. (14.3)]{WM08}. A QND variable can be continuously measured while the measurement at the present time has no degrading effect on the future measurements.  If an input $u$ is decoupled from an output $y$; in other words, the measurement of the output $y$ is not affected by the input $u$,  then we say that the underlying system realizes a quantum BAE  measurement of the output $y$ with respect to the input $u$.  More discussions on these three notions can be found in, e.g., \cite{HMW95}, \cite{TV09}, \cite{TC12}, \cite{NY14}, \cite{GZ15},   \cite{ZGPG18} and references therein.

In this paper, we show that a suitable structural  decomposition of the two-level system \eqref{sys1_a}-\eqref{sys1_b}  enables us to display all of these important concepts in a transparent way. More specifically, the real system matrices in \eqref{A_0}-\eqref{A} for the input-output model \eqref{sys1_a}-\eqref{sys1_b} are parametrized by two row vectors, namely $\alpha$ for the system Hamiltonian $H$ and $\Gamma$  for the coupling operator $L$. In this paper, we aim to find coordinate transformations of the form 
\[
\left[
\begin{array}{c}
\tilde{\sigma}_1\\
\tilde{\sigma}_2\\
\tilde{\sigma}_3
\end{array}
\right]
\equiv\tilde{X}  = T^\top X = T^\top\left[ 
\begin{array}{c}
\sigma _1 \\ 
\sigma _2 \\ 
\sigma _{3}
\end{array}
\right]
\]
 such that the transformed Hamiltonian $H = \alpha T \tilde{X} \equiv \tilde{\alpha}\tilde{X}$ and coupling operator $L = \Gamma T \tilde{X} \equiv \tilde{\Gamma}\tilde{X} $ yield another input-output model which will  facilitate  the study of  properties of two-level systems, such as stationary solutions to the Bloch equation \eqref{eq:jun27_a_2}, DF subspaces, QND variables, and BAE measurements of two-level systems.

In order to perform coordinate transformations on the input-output model  \eqref{sys1_a}-\eqref{sys1_b}, we borrow some well-known  concepts from linear systems theory. Analogous to quantum linear system theory  (\cite{NY14}, \cite{ZGPG18}), we define the observability matrix $\mathcal{O}$ and the controllability matrix $\mathcal{C}$ by
\begin{equation}   \label{obsv_0}
\mathcal{O} \triangleq \left[ 
\begin{array}{c}
C \\ 
CA \\ 
CA^{2}
\end{array}
\right],  \ \ \mathcal{C} \triangleq [ 
\begin{array}{ccc}
B & AB & A^{2}B%
\end{array}%
]. 
\end{equation}
By means of the matrices $\mathcal{O}$ and $\mathcal{C}$ introduced above, we define the following subspaces of $\mathbb{R}^3$:
\begin{eqnarray*}
R_{co} &\triangleq &\mathrm{Range}(\mathcal{C})\cap \mathrm{Range}(O^\top ),
\\
R_{c\bar{o}} &\triangleq &\mathrm{Range}(\mathcal{C})\cap \mathrm{Ker}(\mathcal{O}),
\\
R_{\bar{c}o} &\triangleq &\mathrm{Ker}(\mathcal{C}^\top )\cap \mathrm{Range}(\mathcal{O}^\top ),
 \\
R_{\bar{c}\bar{o}} &\triangleq &\mathrm{Ker}(\mathcal{C}^\top )\cap \mathrm{Ker}(\mathcal{O}). 
\end{eqnarray*}

Finally, we introduce controllability and observability of the input-output model \eqref{sys1_a}-\eqref{sys1_b} of a two-level quantum system in terms of the controllability matrix $\mathcal{C}$ and observability matrix $\mathcal{O}$ defined in Eq. \eqref{obsv_0}. These two notions are again borrowed from linear systems theory (\cite{NY14}, \cite{GZ15}, \cite{ZGPG18}). However, it turns out that they are very useful for structural decompositions of the input-output model \eqref{sys1_a}-\eqref{sys1_b} to be carried out in the next two sections.

\begin{definition}\label{def:ctrb_obsv}
The input-output model \eqref{sys1_a}-\eqref{sys1_b} is said to be controllable if the controllability matrix $\mathcal{C}$ in \eqref{obsv_0} is of full row rank and observable if the observability matrix $\mathcal{O}$ in \eqref{obsv_0} is of full column rank. For simplicity, we say that the matrix pair $(A,B)$ is controllable if the system \eqref{sys1_a}-\eqref{sys1_b}  is controllable, and $(A,C)$ is observable if the system \eqref{sys1_a}-\eqref{sys1_b}  is observable.
\end{definition}

\section{A special case: $\Gamma$ is a non-zero {\it real} row vector} \label{sec:special}

In this case, by Eq. \eqref{c1}, the real row vector $c_{2}=0$. The system matrices in Eqs. \eqref{A_0}-\eqref{A} become 
\begin{subequations}
\begin{align}
&A_0 =  \left[ 
\begin{array}{c}
0 \\ 
0\\
0
\end{array}
\right],  \    
B = 
[ 
\begin{array}{cc}
0 & -\Theta (c_{1})
\end{array}
]  ,  \ C =\left[ 
\begin{array}{c}
c_1 \\ 
0
\end{array}
\right],
 \label{B}
\\
&A 
=
-2\Theta (\alpha )+\frac{1}{2}\left(c_{1}^{\top}c_{1}-c_{1}c_{1}^{\top}I\right) .
\label{A_2}
\end{align}
\end{subequations}

The purpose of this section is to propose suitable coordinate transformations for this special case. The main result is Theorem \ref{thm:c2=0}. To prove Theorem \ref{thm:c2=0}, several lemmas have to be established first.

\begin{lemma}\label{lem:alpha=0}
If $\alpha=0$, then  $\mathrm{Range}(\mathcal{C}) =\mathrm{Ker}(\mathcal{O}) = \mathrm{Ker}(c_1)$.
\end{lemma}
\textbf{Proof.}  If $\alpha =0$, then $c_1 A=0$, and hence $\mathrm{Ker}(\mathcal{O}) =\mathrm{Ker}(c_1)$. Moreover, by $c_1\Theta(c_1)=0$ we have  $\mathrm{Range}(\mathcal{C})=\mathrm{Range}(\Theta(c_1))\subset \mathrm{Ker}(c_1)$.
However,  the dimension of $\mathrm{Range}(\mathcal{C})$ is 2 and we have $\mathrm{Range}(\mathcal{C}) = \mathrm{Ker}(c_1)$. $\blacksquare$

\begin{lemma}\label{lem:ctrl}
If $\alpha \neq 0$ and $c_{1}\neq \mu \alpha $ for all $\mu\in \mbb{R}$,
then $\left( A,B\right) $ is controllable.
\end{lemma}

\textbf{Proof.}  Suppose $\left( A,B\right) $ is not controllable. Then, according to Eqs. \eqref{B}-\eqref{A_2} and Definition \ref{def:ctrb_obsv},  the matrix pair $\left(\Theta(\alpha ), B_{2}\right) $ is not controllable. Thus, there exists a {\it non-zero}
column vector $x\in \mathbb{C}^{3}$ such that 
\begin{equation}\label{xB}
x^\dag B_{2} =-x^\dag \Theta(c_1)=0,  
\end{equation}
and 
\begin{equation} \label{xTheta_alpha}
x^\dag \Theta (\alpha )=\lambda x^\dag  
\end{equation}
for some $\lambda \in \mathbb{C}$.  By (\ref{xB}),  there exist $\mu _{1},\mu _{2}\in \mathbb{R}$ such that 
\begin{equation}\label{temp1a} 
x = \left( \mu _{1}+\imath\mu _{2}\right) c_{1}^{\top}. 
\end{equation}
Substituting Eq. (\ref{temp1a}) into Eq. (\ref{xTheta_alpha}) yields
\begin{equation}\label{eq:mar22_temp1}
\left( \mu _{1}-\imath\mu _{2}\right) c_{1}\Theta (\alpha )=\lambda \left( \mu
_{1}-\imath\mu _{2}\right) c_{1}.
\end{equation}
Noticing that $\mu_1$ and $\mu_2$ cannot be zero simultaneously as $x$ is a non-zero vector,  Eq. \eqref{eq:mar22_temp1} is equivalent to 
\begin{equation} \label{temp2}
c_{1}\Theta (\alpha )=\lambda c_{1}. 
\end{equation}
Post-multiplying both sides of Eq. \eqref{temp2} by $c_{1}^{\top}$ yields  $\lambda c_{1}c_{1}^{\top}=c_{1}\Theta (\alpha )c_1^\top=0$. As a result, $\lambda =0$ and hence $c_{1}\Theta (\alpha )=0$. As $\alpha \neq 0$,  we have
\begin{equation} \label{eq:sept6_1}
c_{1}=\mu \alpha ,
\end{equation}
for some $0\neq \mu \in \mathbb{R}$, which is  a contradiction. Thus, $(A,B) $ must be controllable. $\blacksquare$

\begin{lemma}\label{lem:obsv}
If $c_{1}\neq \mu \alpha $ for all $\mu\in \mbb{R}$,
and the scalar $\alpha c_{1}^{\top}\neq 0$, then $\left( A,C\right) $ is observable.
\end{lemma}

\textbf{Proof.} 
By Eq. \eqref{B}, the second row of the matrix $C$ is zero. To simplify the notation, we identify $C$ with its first row $c_1$.   Suppose $0\neq x\in \mathrm{Ker}(\mathcal{O})$. Then
\begin{subequations}
\begin{eqnarray}
Cx &=&0,  
\label{temp3a} 
\\
C\Theta (\alpha )x &=&0,  
\label{temp3b} 
\\
4C\Theta (\alpha )^{2}x-C\Theta (\alpha )\left( C^{\top}C-CC^{\top}I\right) x &=&0.
\label{temp3c}
\end{eqnarray}
\end{subequations}
Substituting Eqs. (\ref{temp3a})-(\ref{temp3b}) into Eq. (\ref{temp3c}) yields $C\Theta (\alpha )\left( C^{\top}C-CC^{\top}I\right) x=0$.  Consequently, Eq. (\ref{temp3c}) becomes  
$
0=4C\Theta (\alpha )^{2}x=4C\alpha ^{\top}\alpha x-4(\alpha \alpha ^{\top})Cx=4C\alpha ^{\top}\alpha x$. Because $C\alpha ^{\top}=(\alpha C^{\top})^{\top}\neq 0$,  we have
\begin{equation}\label{temp4a}
\alpha x=0.  
\end{equation} 
By Eqs. \eqref{temp3a} and  (\ref{temp4a}),
\begin{equation}\label{temp6a}
x \perp \alpha^\top, \quad x \perp C^\top.  
\end{equation}
Moreover, as $C\Theta (\alpha )$ is a non-zero vector, 
\begin{equation} \label{temp6b}
\left\{ z:C\Theta (\alpha )z=0\right\} =\mathrm{Range}([\alpha^{\top} \ C^{\top}]).
\end{equation}
By Eqs. \eqref{temp3b} and \eqref{temp6b},  $x\in \mathrm{Range}([\alpha^{\top} \ C^{\top}])$.
This, together with Eq. (\ref{temp6a}), implies that $x=0$. Hence, a contradiction is reached.  $\left( A,C\right) $ must be observable. $\blacksquare$

\begin{corollary}\label{cor:c and o}
If  $(A,C)$ is observable, then $(A,B) $ is controllable.
\end{corollary}

\textbf{Proof.}  Assume that $(A,B) $ is not controllable. Then as in the proof of Lemma \ref{lem:ctrl}, there exists a {\it non-zero}
scalar $\mu\in \mathbb{C}$  such that   Eq. \eqref{eq:sept6_1} holds. Thus, $c_1\Theta(\alpha)=0$. Consequently,  $\mathrm{Ker}(\mathcal{O})=\mathrm{Ker}(c_1)$ with dimension 2. That is, $(A,C)$ is not observable. A contradiction is reached.  $\blacksquare$

\begin{lemma}\label{lem:Ker_Im_1}
If $c_1\neq \mu \alpha $ for all $\mu \in \mathbb{R}$, $\alpha\neq0$, 
and $\alpha c_1^{\top}=0$, then  $\mathrm{Ker}(\mathcal{O})=\mathrm{Range}(\alpha ^{\top})$.
\end{lemma}

\textbf{Proof.} First,  it is easy to show that  $\alpha ^{\top}\in \mathrm{Ker}(\mathcal{O})$.   Second,  for any 
$x\in \mathrm{Ker}(\mathcal{O})$,  we have $Cx =0$ and $C\Theta (\alpha )x = 0$. Thus, 
\begin{equation*}
x\in \mathrm{Ker}(C)\cap \mathrm{Ker}(C\Theta (\alpha ))
=
\mathrm{Ker}(C)\cap \mathrm{Range}([\alpha^{\top} \  \ C^{\top}]) ,
\end{equation*}
where Eq. (\ref{temp6b}) has been used. Let $x=x_{1}\alpha ^{\top}+x_{2}C^{\top}$.
Then from $Cx=x_{1}C\alpha ^{\top}+x_{2}CC^{\top}=x_{2}CC^{\top}=0$ we get $x_{2}=0$.
That is, $x=x_{1}\alpha ^{\top} \in \mathrm{Range}(\alpha ^{\top})$.  $\blacksquare$

The following lemma can be established by straightforward algebraic manipulations. Hence its proof is omitted.
\begin{lemma}\label{lem:Ker_Im_2}
If $\alpha =\mu c_1$ for some $\mu \in \mathbb{R}$ (including the case $\alpha =0$ in Lemma \ref{lem:alpha=0}), then $\mathrm{Range}(\mathcal{C}) =  \mathrm{Ker}(\mathcal{O}) = \mathrm{Ker}(c_1)$.
\end{lemma}

Based on the above lemmas, we have the main result of this section.
\begin{theorem}\label{thm:c2=0}
When $\Gamma$ is a non-zero real vector,  we have the following cases and coordinate transformations: 
\begin{description}
\item[(i)] If $c_1\neq \mu \alpha $ for all $\mu \in \mathbb{R}$, and $\alpha c_1^{\top}\neq 0$, then the system is controllable and observable and no coordinate transformation is needed.
\item[(ii)] If $c_1\neq \mu \alpha $ for all $\mu \in \mathbb{R}$, $\alpha\neq0$, and $\alpha c_1^{\top}=0$, then there exists a real orthogonal matrix $T$ which implements the coordinate transformation $\tilde{X} = T^\top X$ with  transformed system matrices
\begin{subequations}
\begin{eqnarray}
\tilde{A} &=&  \left[
\begin{array}{cc}
A_{\rm co} & 0\\
0                & A_{\rm c\bar{o}}
\end{array}
\right],  \  \tilde{A}_0 =  \left[
\begin{array}{l}
    0 \\
     0 \\
     0
\end{array}
\right],
\label{case 2_A} 
\\
\tilde{B} &=&  [0 \ \  \tilde{B}_2],  \  \ 
\tilde{C} =   \left[
\begin{array}{cc}
\tilde{c}_{11} & 0 \\
 0 &0
\end{array}
\right].
 \label{case 2_C} 
\end{eqnarray}
\end{subequations}
\item[(iii)]  If $\alpha =\mu c_1$ for some $\mu \in \mathbb{R}$, then  there exists a real orthogonal matrix $T$ which implements the coordinate transformation $\tilde{X} = T^\top X$ with transformed  system matrices
\begin{subequations}
\begin{eqnarray}
\tilde{A} &=& \left[
\begin{array}{cc}
A_{\rm c\bar{o}} & 0\\
0                & 0
\end{array}
\right],  \  \tilde{A}_0 =  \left[
\begin{array}{l}
    0 \\
     0 \\
     0
\end{array}
\right],
\label{case 3_A} 
\\
\tilde{B} &=&  \left[
\begin{array}{cc}
0 &\tilde{B}_{12}\\
0 & 0
\end{array}
\right], \ 
\tilde{C} =  \left[
\begin{array}{cc}
0&\tilde{c}_{12}  \\
 0  & 0
\end{array}
\right].
\label{case 3_B}  
\end{eqnarray}
\end{subequations}
\end{description}
\end{theorem}

The proof of Theorem \ref{thm:c2=0} is given in {\bf Appendix A}, where the coordinate transformation matrix $T$ is constructed {\it explicitly} in each case.

In Theorem \ref{thm:c2=0}, the coordinate transformation matrix $T\in \mbb{R}^{3\times3}$ is real orthogonal. It is well-known that such a matrix $T$ can be a rotation matrix with $\mathrm{det}(T)=1$ or a Householder matrix with $\mathrm{det}(T)=-1$; see, e.g.,  \cite{NJ85}, \cite{ASH58}. Theorem \ref{thm:c2=0} itself does not tell us which type of real orthogonal matrix the matrix $T$ is. On the other hand, a coordinate transformation has to preserve commutation relations. Specifically, from the coordinate transformation $\tilde{X} = T^\top X$ and Eq. \eqref{eq:aug24_X}, we have $[\tilde{X},~\tilde{X}^T]= 2\imath T^\top\Theta(X)T$. As a result, the commutation relation $[\tilde{X},~\tilde{X}^\top] = 2\imath \Theta({\tilde X})$ for $\tilde{X}$ is equivalent to $\Theta(X) = T \Theta(T^\top X)T^\top$, while the latter holds if the coordinate transformation matrix $T$ satisfies Eq. \eqref{T_CCR} in Lemma \ref{thm:SO3}. We summarize the above discussions in the following theorem.

\begin{theorem} \label{thm:SO3_aug13}
In order that the transformed system is still a quantum system in the sense that the fundamental commutation relations are preserved, the coordinate transformation matrix $T$ in Theorem \ref{thm:c2=0} has to be a rotation matrix.
\end{theorem}

\begin{remark}
According to Theorem \ref{thm:SO3_aug13}, the coordinate transformation matrix $T$ has to be a rotation matrix.  If $T$ constructed in Theorem \ref{thm:c2=0} is indeed  a rotation matrix,  in other words, ${\rm det}(T) = 1$, no further action is required. On the other hand,  if $T$ constructed in Theorem \ref{thm:c2=0} is a Householder matrix,  in other words, ${\rm det}(T) =-1$, we can simply  swap its first and the second columns so that the resulting transformation matrix is a rotation matrix. This swapping of columns in $T$ will not affect the form of the transformed system  in Theorem \ref{thm:c2=0}.
\end{remark}

For quantum linear systems,  there are close relations among controllability, observability and Hurwitz stability, see, e.g., \cite{NY14}, \cite{GZ15}, \cite{ZGPG18}. For the two-level system studied in this paper, from  Corollary \ref{cor:c and o} we see that $(A,B)$ is controllable if $(A,C)$ is observable. In what follows, we give a more detailed discussion. 

We start with a preliminary result.
\begin{lemma}\label{lem:lypa}
Let $A,B,C$ be three real matrices such that $A = B+C$, where $B$ is negative semi-definite and $C$ is skew-symmetric. Then   $(A,B)$ is controllable if and only if $A$ is Hurwitz stable. 
\end{lemma}

\textbf{Proof.}  Firstly, we show the ``only if'' part. Notice that
\begin{equation}\label{eq:lypa}
A+A^\top = 2B\leq0.
\end{equation}
Let $\lambda$ be a scalar and $x$ a nonzero vector such that $Ax = \lambda x$. Then by Eq. \eqref{eq:lypa}, 
\begin{equation}\label{eq:lypa3}
2{\rm Re}(\lambda) x^\dag x =  x^\dag(A+A^\top) x  = 2 x^\dag B x \leq 0.
\end{equation}
As $(A,B)$ is controllable, $x^\dag B \neq 0$. As $B\leq 0$, we actually have  $x^\dag B x \neq 0$. This, together with Eq. \eqref{eq:lypa3}, gives $2{\rm Re}(\lambda) x^\dag x  <0$. Thus, $A$ is Hurwitz stable. 

Next, we establish the ``if'' part by means of contradiction. Assume that $(A,B)$ is not controllable. Then there exist $\lambda$ and $x\neq 0$ such that $Ax = \lambda x$ and $x^\dag B=0$. By \eqref{eq:lypa3},  ${\rm Re}(\lambda)=0$. That is, the matrix $A$ must have  at least one eigenvalue on the imaginary axis. Hence it is not Hurwitz stable. Thus, a contradiction is reached.  $\blacksquare$

The following result is an immediate consequence of Theorem \ref{thm:c2=0} and Lemma \ref{lem:lypa}.

\begin{corollary}\label{thm:A}
\begin{description}
\item[(i)] If $c_1\neq \mu \alpha $ for all $\mu \in \mathbb{R}$, and $\alpha c_1^{\top}\neq 0$, then  the matrix $A$ is Hurwitz stable.
\item[(ii)] If $c_1\neq \mu \alpha $ for all $\mu \in \mathbb{R}$, $\alpha\neq0$, and $\alpha c_1^{\top}=0$,  both $A_{\rm co}$ and $A_{\rm c\bar{o}}$ in Eq. \eqref{case 2_A} are Hurwitz stable.

\item[(iii)] If $\alpha =\mu c_1$ for some $\mu \in \mathbb{R}$,   then $A_{\rm c\bar{o}}$ in Eq. \eqref{case 3_A} is Hurwitz stable. 
\end{description}
\end{corollary}

The proposed structural decomposition manifests various properties of the  input-output model \eqref{sys1_a}-\eqref{sys1_b} of a two-level quantum system, as commented in the following remark.

\begin{remark} \label{rem:special}
We have the following observations.

\begin{enumerate}

\item The coordinate transformations are performed on system variables, not on the system states. Thus, the coordinate transformation matrix $T$ is independent of initial system states.

\item By Corollary \ref{thm:A}, the systems  in Cases (i)-(ii) of Theorem \ref{thm:c2=0}  are Hurwitz stable. Hence  the Bloch equation \eqref{eq:jun27_a_2} has a unique stationary solution $[0 \ \ 0 \ \ 0]^\top$, and the corresponding steady state $\rho_{ss}$ is a completely mixed state, i.e., $\rho_{ss} = \frac{1}{2}I$. 

\item The condition $[H,L]=0$ is assumed in some works, e.g.,  \cite{GBB+08}, \cite{QPG13},  \cite{BP14}, \cite[Sec. 4.4]{MR15}, and \cite{GZP19}. This is exactly the condition in Case (iii) of Theorem \ref{thm:c2=0}. By Corollary \ref{thm:A},  the system in Case (iii) of Theorem \ref{thm:c2=0} is not  Hurwitz stable.   The stationary state is parameterized by $(0,0,\tilde{a}_3(0))$. As a result, if the system starts from an initial state such that
$
\tilde{a}_1(0)=\tilde{a}_2(0)=0, \ \tilde{a}_3(0) = \pm 1,
$
then the stationary state is a pure state $\rho_{ss} = \frac{1}{2}(I\pm \tilde{\sigma}_3)$; otherwise, the steady state is always a mixed state. Moreover, the subspace of states $\tilde{\rho} = \frac{1}{2}(I + a_3 \tilde{\sigma}_3)$ with $a_3 \in [-1, 1]$ forms a DF subspace. In fact, as all such states $\tilde{\rho}$ satisfy $[H, \tilde{\rho}] = \mathcal{L}_{L}(\tilde{\rho})=0$, they are dark states (\cite{SW10}).

\item The structural decomposition given in Theorem \ref{thm:c2=0}  reveals quantum QND variables  and quantum BAE measurements of the two-level system. More specifically,  in Case (i) of Theorem \ref{thm:c2=0}, there is no QND variable; no quantum BAE measurements can be achieved either. Also, in Case (ii) of Theorem \ref{thm:c2=0}, there is no QND variable;  however, the system achieves  quantum BAE measurements  from $W_1$ to $Y_2$. Finally, in Case (iii) of Theorem \ref{thm:c2=0}, $\tilde{\sigma}_3$ is a QND variable as $\tilde{\sigma}_3(t)\equiv \tilde{\sigma}_3(0)$ for all $t\geq 0$; moreover, the system achieves quantum BAE measurements from $W_1$ to $Y_2$ and from $W_2$ to $Y_1$.
\end{enumerate}
\end{remark}

\section{The general case}\label{sec:general}
In this section, we study the general case where $c_2\neq 0$. The main result is Theorem \ref{thm:general}.  First, we present some preliminary lemmas. Although more algebraically complicated,  all of these lemmas can be established in a similar way as in Section \ref{sec:special}. Hence, all their proofs are omitted.

\begin{lemma}\label{lem:ctrb_gen}
If $0\neq c_{1}\neq \mu c_{2}\neq 0$ for all $\mu \in 
\mathbb{R}$, then $(A,B)$ is controllable.
\end{lemma}

\begin{lemma}\label{lem:kr_gene}
If $0\neq c_{1}\neq \mu c_{2}\neq 0$ for all $\mu \in \mathbb{R}$, $\alpha\neq0$,  $\alpha c_{1}^{\top}=0$ and $\alpha c_{2}^{\top}=0$, then  $\mathrm{Ker}(\mathcal{O})=\mathrm{Range}(\alpha ^{\top})$.
\end{lemma}

\begin{lemma}\label{lem:kr_gene_Aug2}
If $0\neq c_{1}\neq \mu c_{2}\neq 0$ for all $\mu \in \mathbb{%
R}$, and $\alpha=0$, then  $\mathrm{Ker}(\mathcal{O})=\mathrm{Ker}\left(C\right)$.
\end{lemma}

\begin{lemma}\label{lem:obsv_gen}
If $0\neq c_{1}\neq \mu c_{2}\neq 0$ for all $\mu \in 
\mathbb{R}$, $\alpha\neq0$, and either $\alpha c_{1}^{\top} \neq 0$ or $\alpha c_{2}^{\top} \neq 0$, then $%
\left( A,C\right) $ is observable.
\end{lemma}

\begin{lemma}\label{lem:sep3_1}
If $0\neq c_{2}=\mu c_{1}\neq 0$ for some $\mu \in \mathbb{R}$, and $\alpha =\nu c_1$ for some $\nu \in \mathbb{R}$ (including the case $\alpha =0$), then $\mathrm{Ker}(\mathcal{O}) =\mathrm{Range}(\mathcal{C}) 
=\mathrm{Ker}(c_1)$.
\end{lemma}

It can be easily shown that Corollary \ref{cor:c and o} proved for the special case in Section \ref{sec:special}  still holds in the general case.

Based on the above results, we have the main result of this section.
\begin{theorem}\label{thm:general}
When $c_1\neq0$ and $c_2\neq0$,  we have the following cases and coordinate transformations. 
\begin{description}
\item[(i)] If $0\neq c_{1}\neq \mu c_{2}\neq 0$ for all $\mu \in \mathbb{R}$, and
either $\alpha c_{1}^{\top} \neq 0$ or $\alpha c_{2}^{\top} \neq 0$, then the system is controllable and observable and no coordinate transformation is needed.
\item[(ii)] If $0\neq c_{1}\neq \mu c_{2}\neq 0$ for all $\mu \in \mathbb{R}$, $\alpha
c_{1}^{\top}=0$ and $\alpha c_{2}^{\top}=0$, then there exists a real orthogonal matrix $T$ which implements the coordinate transformation $\tilde{X}  = T^\top X$
with transformed system matrices
\begin{subequations}
\begin{eqnarray}
\tilde{A} &=& \left[
\begin{array}{cc}
A_{\rm co} & 0\\
0                & A_{\rm c\bar{o}}
\end{array}
\right] ,  \ \ 
\tilde{A}_0  = \left[
\begin{array}{c}
0\\
0\\
A_{0,c\bar{o}}
\end{array}
\right],
\label{sep15_case 2_A} 
\\
\tilde{B} &=& [\tilde{B}_1 \ \ \tilde{B}_2] ,  \ \ 
\tilde{C}=\left[
\begin{array}{cc}
\tilde{c}_{11} & 0 \\
\tilde{c}_{21} &0
\end{array}
\right].
 \label{sep15_case 2_C} 
\end{eqnarray}
\end{subequations}
\item[(iii)]  If $0\neq c_{2}=\mu c_{1}\neq 0$ for some $\mu \in \mathbb{R}$, the coordinate transformation is the same as  that for $c_2=0$ as studied in  Theorem \ref{thm:c2=0}. More specifically,
\begin{description}
\item[(a)] If $c_1\neq \nu \alpha $ for all $\nu \in \mathbb{R}$, and $\alpha c_1^{\top}\neq 0$, then the system is both controllable and observable and no coordinate transformation is needed.
\item[(b)] If $c_1\neq \nu \alpha $ for all $\nu \in \mathbb{R}$, $\alpha\neq0$, and $\alpha c_1^{\top}=0$, then there exists a real orthogonal matrix $T$ which implements the coordinate transformation $\tilde{X} = T^\top X$ with transformed system matrices
\begin{eqnarray*}
\tilde{A} &=& \left[
\begin{array}{cc}
A_{\rm co} & 0\\
0                & A_{\rm c\bar{o}}
\end{array}
\right],  \ \ \tilde{A}_0 =  \left[
\begin{array}{l}
    0 \\
     0 \\
     0
\end{array}
\right],
\\
\tilde{B} &=& [\mu\tilde{B}_2 \ \ -\tilde{B}_2], 
 \ \ 
\tilde{C} =  \left[
\begin{array}{cc}
\tilde{c}_{1} & 0 \\
\mu \tilde{c}_{1} &0
\end{array}
\right].
\end{eqnarray*}
\item[(c)]  If $\alpha =\nu c_1$ for some $\nu \in \mathbb{R}$, then  there exists a real orthogonal matrix $T$ which implements the coordinate transformation $\tilde{X} = T^\top X$ with transformed system matrices
\begin{eqnarray*}
\tilde{A} &=&\left[
\begin{array}{cc}
A_{\rm c\bar{o}} & 0\\
0                & 0
\end{array}
\right],  \ \ \tilde{A}_0 =  \left[
\begin{array}{l}
    0 \\
     0 \\
     0
\end{array}
\right],
\\
\tilde{B} &=& \left[
\begin{array}{cc}
\mu\tilde{B}_{12} &-\tilde{B}_{12}\\
0 & 0
\end{array}
\right],  \ \ 
\tilde{C} =  \left[
\begin{array}{cc}
0&\tilde{c}_{12}  \\
 0  & \mu\tilde{c}_{12}
\end{array}
\right].
\end{eqnarray*}
\end{description}
\end{description}
\end{theorem}

Most part of Theorem \ref{thm:general} can be proved in a similar way to that for Theorem \ref{thm:c2=0}, except for  the special case $\alpha=0$ in Case (ii). Hence, we give the proof of this special case in {\bf Appendix B}, while omitting all of the other proofs.

We end this section with a final remark.

\begin{remark} \label{rem:general}
We have the following observations.

\begin{enumerate}
\item In both the special and general cases, the Bloch equation \eqref{eq:jun27_a_2}  always has a stationary solution, thus the Lindblad master equation \eqref{eq:master} always has a steady state.  The existence of steady states of the Lindblad master equation for general $n$-level systems has been proved in \cite[Proposition 1]{SW10}, where the mathematical tools such as Brouwer's fixed point theorem and Cantor's intersection theorem are used.  We have not seen in the literature a simpler proof for the $n=2$ case. However,   Theorems \ref{thm:c2=0} and \ref{thm:general} show that the proposed structural decompositions manifest the existence of steady states transparently.

\item The structural decomposition  shows clearly which part of the system is affected by the input noises $W_1$ and $W_2$, and which part is not. This enables us to find quantum QND variables of the two-level system. Moreover, it allows us to study quantum BAE measurements. More specifically,  In Cases (i), (ii), (iii(a)) and  (iii(b))  of Theorem \ref{thm:general},  there is no QND variable. Moreover, no BAE measurement can be realized in these cases. Also, in   Case (iii(c)) of Theorem \ref{thm:general}, $\tilde{\sigma}_3$ is a QND variable as $\tilde{\sigma}_3(t)\equiv \tilde{\sigma}_3(0)$ for all $t\geq 0$. Furthermore in this case, the system achieves BAE measurement from $W_1$ to $Y_2$ and from $W_2$ to $Y_1$.

\item According to Corollary \ref{cor:c and o}, observability leads to controllability. Moreover, by  Lemma \ref{lem:lypa}, controllability is equivalent to Hurwitz stability (which implies a unique stationary solution to the Bloch equation \eqref{eq:jun27_a_2}).  Nonetheless, according to Theorems  \ref{thm:c2=0} and \ref{thm:general}, a two-level system may be uncontrollable and observable, which gives rise to QND variables. Moreover, in some cases, a two-level system may allow quantum BAE measurements; see item (2) above.

\item In Section \ref{sec:example}, we will use the physical system in \cite{WW01} to show that measurement-based  feedback control can change the system Hamiltonian $H$ and the Lindblad coupling operator $L$, thus affecting the structural decomposition and properties of a two-level quantum system.

\end{enumerate}
\end{remark}

\section{An example} \label{sec:example}
In \cite{WW01}, the authors showed how to use measurement-based feedback control to stabilize pure states of a two-level atom. Before feedback, the  $H$ and $L$ are respectively $H = \alpha_2 \sigma_2$ and $L = \Gamma X$, where  $\Gamma = \frac{\sqrt{\gamma}}{2} [1 \  -\imath \ 0]$. Now, we consider feedback with unit detection efficiency \cite[Sec. III]{WW01}. However, instead of measuring $Y_1$ in Eq. \eqref{sys1_b} only,  we choose to measure  $\cos(\phi)Y_1 + \sin(\phi)Y_2$, where $\phi\in[-\pi, \pi]$.  This amounts to simply changing the quadrature measured by the Homodyne detector. We use the network theory developed in (\cite{GJ09}) to derive the Hamiltonian $H_{cl}$ and coupling operator $L_{cl}$ for the closed-loop system. We have
\beq \label{G_cl}
H_{cl} = H+\frac{\sqrt{\gamma}\lambda \sin(\phi)}{2}(\sigma_3-I), \ \ L_{cl} =  L-\imath \lambda \sigma_2.
\eeq
(Notice that the term $-\frac{\sqrt{\gamma}\lambda\sin(\phi)}{2}I$ in  $H_{cl}$ can be safely ignored.) Accordingly,
\begin{subequations}
\beqn
\alpha_{cl} &=& 
\left[
\begin{array}{ccc}
0 & \ \ \ \  \  \alpha_2 \ \ \ \ \ \  \ \ \ \ \   \ \  \   & \frac{\sqrt{\gamma}\lambda}{2}\sin(\phi)
\end{array}
\right], 
\label{alpha_cl}
\\
c_{1,cl} &=&
\left[
\begin{array}{ccc}
\sqrt{\gamma} & \ \   2\lambda \sin(\phi) \ \ \ \ \ \ \ \     \ \ \ \ \ \ \ \ \ \     & 0
\end{array}
\right],
\label{c1_cl}
\\
c_{2,cl} &=& 
\left[
\begin{array}{ccc}
0& \ \ \ \ \  -(\sqrt{\gamma} + 2\lambda\cos(\phi)) & \  \ \ \  0
\end{array}
\right].
\label{c2_cl}
\eeqn
\end{subequations}
With these parameters, we have
\begin{align*}
&A_{cl}
\\
 =&-2\left[
\begin{array}{ccc}
  \lambda ^2+ \sqrt{\gamma }\lambda \cos (\phi )  +\frac{\gamma }{4} & 0 &  -\alpha_2  \\
 - \sqrt{\gamma } \lambda  \sin (\phi ) & \frac{\gamma }{4} & 0 \\
  \alpha_2  & 0 &  \lambda ^2+ \sqrt{\gamma }\lambda \cos (\phi )  +\frac{\gamma}{2}  \\
\end{array}
\right],
\\
&A_{0,cl}
 = -2
\left[
\begin{array}{c}
0\\
0\\
 \sqrt{\gamma } \lambda  \cos (\phi )+\frac{\gamma}{2}
\end{array}
\right].
\end{align*}
Consequently, the Bloch equation after feedback is
\beq \label{eq:jun27_a_3}
\dot{a}_{cl} = A_{cl} a_{cl} +A_{0,cl},
\eeq
whose stationary solution is
\begin{subequations}
\begin{align*}
a_{cl,1} =& -\frac{\alpha_2(\frac{\gamma}{2}+\sqrt{\gamma}\lambda\cos(\phi))}{\Upsilon},
\\
a_{cl,2} =&-\frac{4\alpha_2\lambda  \sin(\phi) (\frac{\gamma}{2}+\sqrt{\gamma}\lambda\cos(\phi))}{\Upsilon\sqrt{\gamma}},
\\
a_{cl,3} =& -\frac{(\frac{\gamma}{2}+\sqrt{\gamma}\lambda\cos(\phi))(\frac{\gamma}{4}+\sqrt{\gamma}\lambda\cos(\phi) + \lambda^2)}{\Upsilon}
\end{align*}
\end{subequations}
with $\Upsilon = \alpha_2^2+(\frac{\gamma}{4}+\sqrt{\gamma}\lambda\cos(\phi) + \lambda^2)(\frac{\gamma}{2}+\sqrt{\gamma}\lambda\cos(\phi) + \lambda^2)$. 

In what follows, we discuss several scenarios to illustrate Cases (ii)-(iii) of Theorem \ref{thm:c2=0}  and Cases (i)-(ii) of Theorem \ref{thm:general}. The closed-loop systems in Scenarios 1 and 4 are naturally in the form of those in Theorems \ref{thm:c2=0} and \ref{thm:general}. However, in Scenarios 2 and 3, the coordinate transformation matrix $T$ ie needed to be constructed to transform the closed-loop system to those of Theorems \ref{thm:c2=0} and \ref{thm:general}. In all of  these scenarios, stationary solutions to Eq. \eqref{eq:jun27_a_3} are given. Moreover, it is shown that quantum BAE measurements can be realized by the system in Scenario 3.

{\bf Scenario 1.} 
If $\lambda=0$ and $\gamma\neq0$, in other words, there is no feedback but the two-level atom is damped, by Eqs. \eqref{alpha_cl}-\eqref{c2_cl}, we have
\[
\alpha_{cl}= [0 \ \alpha_2 \ 0], \ 
c_{1,cl} =  [\sqrt{\gamma} \ 0 \ 0], \  c_{2,cl} = [0 \ -\sqrt{\gamma}\ 0].
\]
Firstly, assume $\alpha_2\neq0$. This is Case (i) of Theorem \ref{thm:general}. As the matrix $A_{cl}$ is Hurwitz stable and $A_{0,cl}\neq 0$, there is a unique stationary solution to Eq. \eqref{eq:jun27_a_3}
\beqm
a_{cl,1} = -\frac{4\alpha_2 \gamma}{\gamma^2+8\alpha_2^2}, \ a_{cl,2}=0,\ a_{cl,3} = -\frac{\gamma^2}{\gamma^2+8\alpha_2^2},
\eeqm
which is the same as that in \cite[Eqs. (2.3)-(2.5)]{WW01}. As $a_{cl,1}^2+a_{cl,2}^2+a_{cl,3}^2<1$, we end up with a mixed state. Secondly, assume $\alpha_2=0$.  In other words, the atom is driven by a vacuum field. This is Case (ii) of Theorem \ref{thm:general}.  The stationary solution becomes $(0,0,-1)$. In this case,  $\rho = \ket{1}\bra{1}$, namely the ground state of the two-level atom, which is certainly stable. 

{\bf Scenario 2.} If $\phi=0$ which is the feedback scheme considered in \cite{WW01}, then $a_{cl,2}=0$. Thus, a pure state 
\beq\label{eq:theta}
\ket{\theta} = \cos\frac{\theta}{2}\ket{0}+\sin\frac{\theta}{2}\ket{1} ,  \ (-\frac{\pi}{2}\leq \phi\leq\frac{\pi}{2})
\eeq
 can be parameterized by  $a_{cl,1} =\sin \theta$ and $a_{cl,3} =\cos \theta$. The stationary solution \eqref{eq:theta} to Eq. \eqref{eq:jun27_a_3}  requires that
\beq \label{lambda_alpha}
\lambda = -\frac{\sqrt{\gamma}}{2} (1+\cos\theta), \ \alpha_2 = \frac{\gamma}{4}\sin\theta\cos\theta,
\eeq
which are \cite[Eqs. (3.8)-(3.9)]{WW01}. Moreover, if $\lambda = -\frac{\sqrt{\gamma}}{2}$, then $\theta = \pm \frac{\pi}{2}$ and $\alpha_2 =0$. As a result,  $c_{1,cl} = [\sqrt{\gamma} \ 0 \ 0]$, and $ c_{2,cl} = \alpha_{cl} = [0 \ 0 \ 0]$. This is the Case (iii) of Theorem \ref{thm:c2=0}. The matrix $A_{cl}$  has a zero eigenvalue. The stationary solutions  to Eq. \eqref{eq:jun27_a_3} are $a_{cl,1}=\pm1 , \ a_{cl,2}=a_{cl,3}=0$. The corresponding steady-state states $\frac{1}{\sqrt{2}}(\ket{0}\pm \ket{1})$  are not asymptotically stable. According to Theorem \ref{thm:c2=0}, a coordinate transformation matrix $T$ can be constructed to transform the closed-loop system to that in Case (iii) of Theorem 1. Due to page limit, the construction of such a matrix is omitted. However, a coordinate transformation matrix $T$ is explicitly given in Scenario 3 below.

{\bf Scenario 3.}  If $\lambda = -\frac{\sqrt{\gamma}}{2}$, $\phi=0$, and $\alpha_2\neq0$, then we get Case (ii) of Theorem \ref{thm:c2=0}. In this case, $A_{0,cl}$ is a zero column vector and $A_{cl}$ is Hurwitz stable. Consequently,  the steady state $\rho_{ss} = \frac{1}{2}I$, which is a completely mixed state. Let $\alpha_2=1$ and $\gamma = 4$. Under the coordinate transformation matrix
\[
T = \left[
\begin{array}{ccc}
 -1 & 0 & 0 \\
 0 & 0 & 1 \\
 0 & 1 & 0 \\
\end{array}
\right],
\]
the system matrices of the {\it transformed} closed-loop system are
\beqnm
&&\tilde{A}_{cl} = \left[
\begin{array}{ccc}
 0 & -2 & 0 \\
 2 & -2 & 0 \\
 0 & 0 & -2 \\
\end{array}
\right], \ 
\tilde{C}_{cl}=
\left[
\begin{array}{ccc}
-2 & 0  & 0\\
 0 & 0 & 0 \\
\end{array}
\right], \   \tilde{A}_{0,cl}
 =
\left[
\begin{array}{c}
0\\
0\\
0
\end{array}
\right],
\\
&& \tilde{B}_{cl} = [0  \ \ \tilde{B}_{2,cl}] = \left[
\begin{array}{cccccc}
 0 & 0 & 0 & 0 & 0 & 0 \\
 0 & 0 & 0 & 0 & 0 & -\sqrt{\gamma } \\
 0 & 0 & 0 & 0 & \sqrt{\gamma } & 0 \\
\end{array}
\right].
\eeqnm
Clearly, the system realizes  quantum BAE measurements  from $W_1$ to $Y_2$, as predicted in Case (ii) of Theorem \ref{thm:c2=0}.

{\bf Scenario 4.}  If $\alpha_2 =0$, $\sin(\phi)\neq 0$, $\lambda\neq0$, $\gamma\neq0$, and $\sqrt{\gamma} + 2\lambda\cos(\phi)\neq 0$, this is Case (ii) of Theorem  \ref{thm:general}. As the matrix $A_{cl}$ is Hurwitz stable and $A_{0,cl}\neq 0$, there is a unique stationary solution to Eq. \eqref{eq:jun27_a_3}, which is
\beqm
a_{cl,1} = 0, \ a_{cl,2}=0,\ a_{cl,3} =-\frac{\gamma +2 \sqrt{\gamma } \lambda  \cos (\phi )}{\gamma +2 \sqrt{\gamma } \lambda  \cos (\phi )+2 \lambda ^2},
\eeqm
which leads to a mixed steady state.

%

\section{Conclusion} \label{sec:con}
In this paper, we have constructed coordinate transformations for an input-output model of two-level quantum systems in the Heisenberg picture. These structural decompositions enable us to investigate many properties of two-level systems, for example stationary solutions to the Lindblad master equation, quantum decoherence-free subspaces, quantum non-demolition variables, as well as quantum back-action evading measurements. The generalization of these results to general $n$-level systems is one possible area of future research.

\textbf{Acknowledgements}
The authors wish to thank Pierre Rouchon and Haidong Yuan for helpful discussions.

\noindent {\bf Appendix A.}

{\bf Proof of Theorem \ref{thm:c2=0}.} (i) In this case,  by Lemma \ref{lem:ctrl}, the
system is controllable. By Lemma \ref{lem:obsv}, the system is observable.
Therefore, $R_{c\bar{o}}=R_{\bar{c}\bar{o}}=R_{\bar{c}o}=\left\{ 0\right\}$  and $R_{co}=\mathbb{R}^{3}$. 
No coordinate transformation is needed. 

(ii) In this case,  by Lemma \ref{lem:ctrl}, the system is controllable. By Lemma \ref{lem:Ker_Im_1}, 
$\mathrm{Ker}(\mathcal{O})=\mathrm{Range}(\alpha ^{\top})$. We have $R_{co} =\mathrm{Range}(\alpha ^{\top})^{\bot }=\mathrm{Ker}(\alpha)$, $R_{c\bar{o}}=\mathrm{Range}(\alpha ^{\top})$, and  $R_{\bar{c}o}= R_{\bar{c}\bar{o}} =\{0\}$. Performing a singular value decomposition (SVD) on $\Theta (\alpha )$, we get $\Theta (\alpha )=U\Lambda V^{\top}$,  where 
\begin{equation*}
U=\left[ 
\begin{array}{cc}
U_1 & U_{2}
\end{array}
\right] , \ \Lambda =\left[ 
\begin{array}{cc}
\Lambda _1 & 0 \\ 
0 & 0
\end{array}
\right] , \ V=\left[ 
\begin{array}{cc}
V_1 & V_{2}
\end{array}
\right] .
\end{equation*}
Define a real orthogonal matrix $T$ to be 
\begin{equation}
T\equiv [T_{co}  \ | \  T_{c\bar{o}}] \triangleq 
[ U_1 \ | \ U_2].
\end{equation}
Then
\begin{subequations}
\begin{eqnarray}
&&\mathrm{Range}(T_{co}) =\mathrm{Range}(\Theta (\alpha )),
\label{ImT1}
\\
&&\mathrm{Range}(T_{c\bar{o}})  =\mathrm{Ker}(\alpha )^{\bot }.
\label{ImT2}
\end{eqnarray}
\end{subequations}
By Eqs.  \eqref{ImT1}-\eqref{ImT2}, it can be shown that $\Theta(\alpha)T_{c\bar{o}}=0$,  $T_{c\bar{o}}^\top \Theta(c_1) \neq 0$, $T_{co}^\top \Theta(c_1) \neq 0$, and $T_{co}^\top \Theta(c_1)^2 T_{c\bar{o}}  = 0$. Consequently,
\begin{equation} \label{sep1_temp9}
T^\top B_2 B_2^\top  T=
\left[ 
\begin{array}{cc}
-T_{co}^\top \Theta(c_1)^2 T_{co}  & 0\\
0 & c_1 c_1^\top 
\end{array}
\right],
\end{equation}
\begin{eqnarray}
&&\tilde{A} =T^{\top}AT
\label{sep1_temp10}
\\
&=&
\left[
\begin{array}{cc}
-2T_{co}^\top \Theta(\alpha)T_{co}+ \frac{1}{2}T_{co}^\top \Theta(c_1)^2 T_{co} & 0 \\
0 &  -\frac{1}{2} c_1 c_1^\top 
\end{array}
\right],
 \nonumber
\end{eqnarray}
which is of the form \eqref{case 2_A}. Moreover, 
\begin{eqnarray}
\tilde{B} &=&T^{\top}B\left[
\begin{array}{cc}
T & 0\\
0 & T
\end{array}
\right] =  -\left[
\begin{array}{cc}
0 &T_{co}^\top \Theta(c_1)\\\
0 & T_{c\bar{o}}^\top \Theta(c_1)
\end{array}
\right],
\label{sep1_temp12}
\end{eqnarray}
and
\begin{equation} \label{sep1_temp11}
CT = \left[
\begin{array}{cc}
c_1 T_{co} & c_1 T_{c\bar{o}}\\
0 &0
\end{array}
\right] =
\left[
\begin{array}{cc}
c_1 T_{co} & 0\\
0 &0
\end{array}
\right],
\end{equation}
which are of the forms \eqref{case 2_C}. 
(Notice that $c_1 T_{co}\neq 0$. Otherwise, $\mathrm{Range}(T_{co}) =  \mathrm{Ker}(\alpha) =  \mathrm{Ker}(c_1)$, which means that $\alpha$ and $c_1$ must be proportional.) 

(iii) In this case,  by Lemma \ref{lem:Ker_Im_2},  $R_{co}  = R_{\bar{c}\bar{o}} =\{0\}$,  $R_{c\bar{o}}  = \mathrm{Ker}(c_1)$, and $R_{\bar{c}o} =\mathrm{Range}(c_1^\top)$. Performing an SVD on $\Theta (c_1 )$, we get $\Theta (c_1)=U\Lambda V^{\top}$,  where 
\[
U=\left[ 
\begin{array}{cc}
U_1 & U_{2}
\end{array}
\right] , \ \Lambda =\left[ 
\begin{array}{cc}
\Lambda _1 & 0 \\ 
0 & 0
\end{array}
\right] , \ V=\left[ 
\begin{array}{cc}
V_1 & V_{2}
\end{array}
\right] .
\]
Define a real orthogonal matrix $T$ to be 
\begin{equation}
T\equiv [T_{c\bar{o}}  \ | \  T_{\bar{c}o}] \triangleq 
[ U_1 \ | \ U_2].
\end{equation}
Then
\begin{subequations}
\begin{eqnarray}
\mathrm{Range}(T_{c\bar{o}}) = \mathrm{Range}(\Theta(c_1)),
\label{sept2_temp1a}
\\
\mathrm{Range}(T_{\bar{c}o})= \mathrm{Range}(c_1^\top).
\label{sept2_temp1b}
\end{eqnarray}
\end{subequations}
By $\alpha = \mu c_1$ and Eq. \eqref{sept2_temp1b} we have $\Theta(\alpha)T_{\bar{c}o}=0$. Thus, \begin{equation} \label{sep2_temp3}
T^\top \Theta(\alpha) T = \left[
\begin{array}{cc}
T_{c\bar{o}}^\top \Theta(\alpha)T_{c\bar{o}} & 0 \\
0 &  0 
\end{array}
\right].
\end{equation}
On the other hand,
\begin{equation} \label{sep2_temp4}
T^\top B_2 
=
-\left[ 
\begin{array}{c}
T_{c\bar{o}}^\top \Theta(c_1)\\
0
\end{array}
\right].
\end{equation}
By Eqs. \eqref{sep2_temp3} and \eqref{sep2_temp4} we get
\begin{eqnarray*}
\tilde{A} =T^{\top}AT
=
\left[
\begin{array}{cc}
-2T_{c\bar{o}}^\top \Theta(\alpha)T_{c\bar{o}}+ \frac{1}{2}T_{c\bar{o}}^\top \Theta(c_1)^2T_{c\bar{o}}  & 0 \\
0 &  0
\end{array}
\right].
\end{eqnarray*}
As a result, the matrix $\tilde{A}$ is of the form \eqref{case 3_A}.  Also, by Eq. \eqref{sep2_temp4},
\begin{eqnarray*}
\tilde{B} &=&T^{\top}B\left[
\begin{array}{cc}
T & 0\\
0 & T
\end{array}
\right] =  \left[
\begin{array}{cc}
0 &-T_{c\bar{o}}^\top \Theta(c_1)\\\
0 & 0
\end{array}
\right].
\label{sep2_temp6}
\end{eqnarray*}
As a result, the matrix $\tilde{B}$ is of the form \eqref{case 3_B}. Finally,  by Eq. \eqref{sept2_temp1a},
\begin{equation*} \label{sep2_temp7}
\tilde{C}=CT = \left[
\begin{array}{cc}
c_1 T_{c\bar{o}} & c_1 T_{\bar{c}o}\\
0 &0
\end{array}
\right] =
\left[
\begin{array}{cc}
0 & c_1 T_{\bar{c}o}\\
0 &0
\end{array}
\right].
\end{equation*}
(Notice that $c_1 T_{\bar{c}o}\neq0$ as $C\neq0$.)  As a result, the matrix $\tilde{C}$ is of the form \eqref{case 3_B}.  Thus, the proof is completed. 
$\blacksquare$

\noindent {\bf Appendix B.}

{\bf Proof of  the case $\alpha =0$ in Case (ii) of Theorem \ref{thm:general}}.  
In this case,  $R_{co} =  \mathrm{Range}(\mathcal{O}^\top)=  \mathrm{Range}(C^\top)$, $R_{c\bar{o}} = \mathrm{Ker}(\mathcal{O})=\mathrm{Ker}\left(C\right)$, and   $R_{\bar{c}o}  =R_{\bar{c}\bar{o}} = \{ 0\}$.
Performing an SVD on $C^\top$, we get $C^\top=U\Lambda V^{\top}$, 
where 
\[
U=\left[ 
\begin{array}{cc}
U_1 & U_{2}
\end{array}
\right] , \ \Lambda =\left[ 
\begin{array}{cc}
\Lambda _1 & 0 \\ 
0 & 0
\end{array}
\right] , \ V=\left[ 
\begin{array}{cc}
V_1 & V_{2}
\end{array}
\right] .
\]
Define a real orthogonal matrix $T$ to be
\begin{equation} 
T\equiv [T_{co} \ | \  T_{c\bar{o}}] \triangleq 
[ U_1 \ | \ U_2].
\end{equation}
Then
\begin{subequations}
\begin{eqnarray}
\mathrm{Range}(T_{co}) = \mathrm{Range}(C^\top),
\label{sept2_temp1a_aug6}
\\
\mathrm{Range}(T_{c\bar{o}})=\mathrm{Ker}\left(C\right).
\label{sept2_temp1b_aug6}
\end{eqnarray}
\end{subequations}
Let $T_{co} = x_1 c_1^\top + x_2 c_2^\top$. By Eqs. \eqref{sept2_temp1a_aug6}-\eqref{sept2_temp1b_aug6} we have
\beqnm
&&T_{co}^\top \Theta(c_1)^2  T_{c\bar{o}} =(x_1 c_1+x_2 c_2) \Theta(c_1)\Theta(c_1)T_{c\bar{o}} 
\\
&=&
-x_2 c_2\Theta(c_1)\Theta(T_{c\bar{o}} )c_1^\top
=
-x_2 c_2(T_{c\bar{o}}  c_1-c_1 T_{c\bar{o}}  I)c_1^\top 
\\
&=& 0.
\eeqnm
  Similarly, it can be easily shown that $T_{co}^\top \Theta(c_2)^2  T_{c\bar{o}}=0$. As a result,
\[
\tilde{A} =T^{\top}AT
=
\left[
\begin{array}{cc}
T_{co}^\top A T_{co} & 0  \\
0 & T_{c\bar{o}}^\top A T_{c\bar{o}}
\end{array}
\right],
\]
which is of the form \eqref{sep15_case 2_A}.  Moreover, as the system is controllable, it is Hurwitz stable. Therefore  the scalar $ A_{c\bar{o}}\triangleq T_{c\bar{o}}^\top A T_{c\bar{o}} \neq0$. 
As $CT_{c\bar{o}}=0$,  we have
\[
\tilde{C} = CT =  \left[
\begin{array}{cc}
\tilde{c}_{11} & 0 \\
\tilde{c}_{21} &0
\end{array}
\right],
\]
which is of the form \eqref{sep15_case 2_C}. 
Finally, it can be readily shown that
\[
\tilde{A}_{0}=T^\top A_0 
=
\left[
\begin{array}{c}
0\\
T_{c\bar{o}}^\top \Theta(c_2) c_1^\top
\end{array}
\right] \equiv   \left[
\begin{array}{c}
0\\
0\\
A_{0,c\bar{o}}
\end{array}
\right],
\]
which is of the form \eqref{sep15_case 2_A}. Clearly,  the real scalar $\tilde{A}_{0,c\bar{o}} \neq 0$ as $A_0\neq0$. Thus, the proof is completed. 
$\blacksquare$



\end{document}